\documentclass[aps,prb,twocolumn,showpacs,superscriptaddress]{revtex4-2}  
\usepackage{float}
\usepackage{bbm}
\usepackage{epsfig}
\usepackage{epstopdf}
\usepackage{graphicx}
\usepackage{amsmath,amssymb}
\usepackage{amsmath,bm}
\usepackage{physics}
\usepackage{color}
\usepackage{hyperref}
\usepackage{lineno,blindtext}
\usepackage{siunitx, booktabs}
\usepackage{diagbox, eqparbox, hhline}
\usepackage{soul}
\usepackage{xcolor}
\usepackage{eufrak}
\usepackage{dsfont}
\usepackage[normalem]{ulem}

\newcolumntype{P}[1]{>{\centering\arraybackslash}p{#1}}

\begin{document}

\title{Non-Hermitian and Liouvillian skin effects in magnetic systems} 
\begin{abstract}

The non-Hermitian skin effect (NHSE) has emerged as a hallmark of non-Hermitian physics, with far-reaching implications for transport, topology, and sensing. While recent works have uncovered the NHSE in magnetic systems, these analyses rely on effective non-Hermitian  Hamiltonians, thereby leaving open critical questions regarding their applicability and predictive power in experimentally feasible platforms.
Here, we address this gap by exploring both the non-Hermitian and Liouvillian dynamics of a spin chain coupled to a shared bosonic reservoir. 
We identify the parameter regime in which these frameworks yield congruent predictions, while showing that the non-Hermitian approach fails to capture essential dynamical features — such as relevant timescales and conditions for experimental observability.
Our analysis also reveals that the NHSE stems from the interplay between chiral spin couplings and reciprocal nonlocal dissipation—two interactions that can naturally occur in magnetic crystals and  be easily engineered in magnetic heterostructures.
 Focusing on a concrete example of such heterostructures, we establish an explicit connection between their Landau–Lifshitz–Gilbert (LLG) dynamics and our microscopic model, providing  a tangible route toward realizing the NHSE in an experimentally relevant spintronics setup.

\end{abstract}
\author{Xin Li}
\email{licqp@bc.edu}
\affiliation{Department of Physics, Boston College, 140 Commonwealth Avenue Chestnut Hill, MA 02467, USA}
\author{Mohamed Al Begaowe}
\affiliation{Department of Physics, Boston College, 140 Commonwealth Avenue Chestnut Hill, MA 02467, USA}
\author{Shu Zhang}
\affiliation{Collective Dynamics and Quantum Transport Unit, Okinawa Institute of Science and Technology Graduate University, Okinawa 904-0495, Japan}
\affiliation{Max Planck Institute for the Physics of Complex Systems, 01187 Dresden, Germany}
\author{Benedetta Flebus}
\affiliation{Department of Physics, Boston College, 140 Commonwealth Avenue Chestnut Hill, MA 02467, USA}

\maketitle

The interplay between non-Hermitian physics and many-body dynamics has opened new avenues for understanding and engineering transport phenomena in open quantum systems~\cite{ashida2020non,bender2007making,el2018non,bergholtz2021exceptional}. A striking manifestation of non-Hermitian behavior is the NHSE, wherein an extensive number of eigenstates accumulate at system boundaries, defying conventional bulk-boundary correspondence~\cite{bergholtz2021exceptional,zhang2022review,song2019non,longhi2020unraveling}. Originally proposed in the context of non-Hermitian tight-binding models, the NHSE has since been studied and enginereed across a wide range of platforms—including photonic~\cite{xiao2020non,weidemann2020topological}, acoustic~\cite{zhang2021acoustic}, electrical~\cite{liu2021non,hofmann2020reciprocal}, mechanical~\cite{ghatak2020observation}, and cold atom systems~\cite{liang2022dynamic}—and proposed as a mechanism for enhanced quantum sensing and signal amplification~\cite{Bao2022,Budich2020,Koch2022,el2015optical,Koutserimpas2018,Wang2022}.

Magnetic systems have recently emerged as a fertile ground for exploring non-Hermitian phenomena, where spin dynamics are governed not only by dissipative processes—such as spin pumping and Gilbert damping 
 ~\cite{hickey2009origin,tserkovnyak2002enhanced,tserkovnyak2002spin}—but also by coherent, nontrivial interactions, including chiral exchange and spin-orbit coupling~\cite{li2022multitude,gunnink2022nonlinear,mcclarty2019non,flebus2020non,yu2022giant,yuan2022master, deng2022non,hurst2022non,deng2023exceptional,lu2021magnetic,zhang2020dynamic,zou2024,flebus2023recent}. Motivated by this rich landscape, recent theoretical efforts have proposed realizations of the NHSE in magnetic media~\cite{deng2022non,yu2022giant}.
However, these works have relied on analyses of effective non-Hermitian  Hamiltonians, raising fundamental questions about their validity 
in studying platforms with
unconditioned dynamics governed by a full Lindblad master equation. In particular, it remains unclear under what conditions the non-Hermitian formalism accurately captures nonreciprocal transport and boundary accumulation in 
an extended spin system interfacing an equilibrium reservoir,  where quantum jumps and detailed balance constraints may play a critical role.

In this work, we address this open question by analyzing a spin chain coupled to a shared bosonic reservoir, comparing its effective non-Hermitian description with the full Liouvillian dynamics. We show that, in the dilute magnon limit and in the absence of external spin injection, the two frameworks become formally equivalent—placing the NHSE and its Liouvillian counterpart on equal footing. Nevertheless, we find that even within this regime, the non-Hermitian approach fails to capture essential aspects of the dynamics, including the relevant timescales and the conditions under which nonreciprocal magnon transport becomes experimentally observable~\cite{Haga2021,lee2023anomalously,Begg-hanai-2024}.

Building on insights from reservoir-engineered nonreciprocity in open quantum systems~\cite{metelmann2015nonreciprocal,metelmann2017nonreciprocal,li2024cooperative}, we also identify the minimal ingredients required to realize a magnetic NHSE: the interplay between coherent chiral interactions—such as Dzyaloshinskii–Moriya interactions (DMI)—and nonlocal dissipation. This finding highlights multiple viable routes to engineer or harness the NHSE in experimentally accessible platforms. Focusing on magnetic heterostructures—where nonmagnetic elements have been shown to mediate both chiral coherent and nonlocal dissipative couplings~\cite{cheng2020nonlocal,ma2016interfacial,skoropata2020interfacial}—we uncover the emergence of the NHSE in their LLG magnetization dynamics, providing a pathway to a concrete experimental realization of non-Hermitian spin transport in magnetic systems.

\begin{figure*}
    \centering    \includegraphics[width=0.9\linewidth]{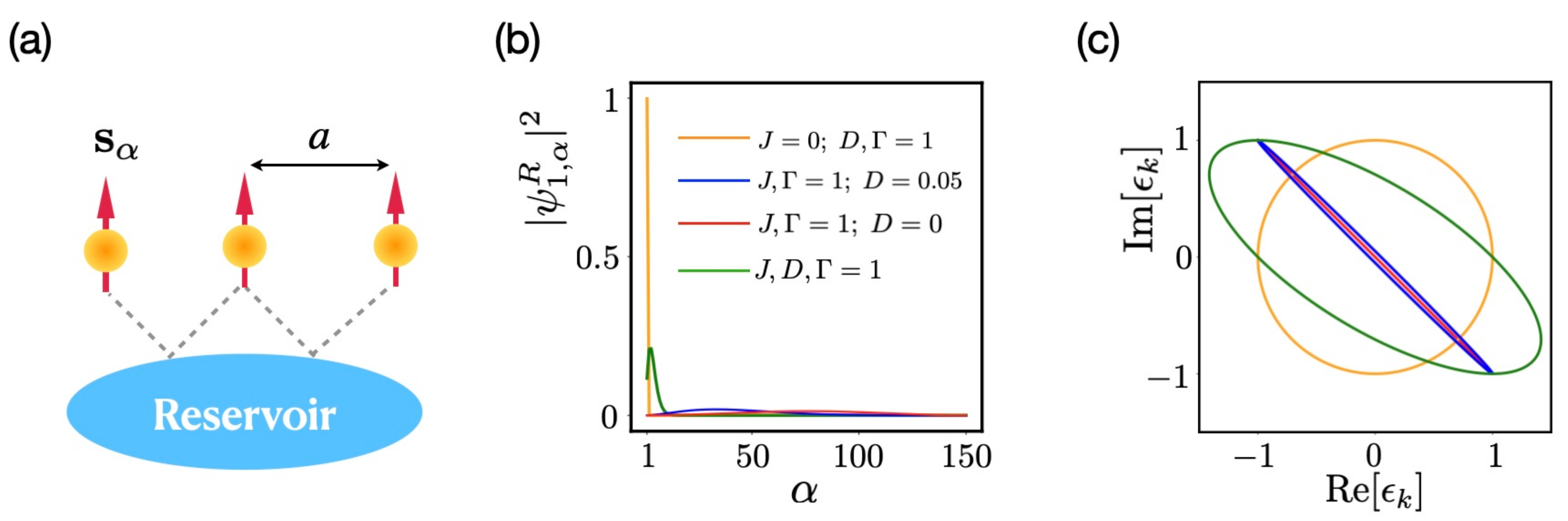}
  \caption{(a) A 1d array of spins interacting with a shared reservoir. $\mathbf{s}_{\alpha}$ denotes the spin at the $\alpha$th lattice site and $a$ is the lattice constant. 
(b) Spatial profile of the density of the first right eigenmode $\psi^{R}_{1,\alpha}$ under open boundary conditions for various sets of system parameters. (c) Corresponding eigenenergy spectra (normalized by $2s$) under periodic boundary conditions.}
\label{fig:master}
\end{figure*}

\section{Model} 
The model under consideration, sketched in Fig.~\ref{fig:master}(a), is a one-dimensional (1$d$) chain of spins subjected to a magnetic field along the $z$ direction.  The spins interact with a common stationary bath according to $\mathcal{H}_\text{int}= \sum_{\alpha} \hat{s}^{+}_{\alpha} \otimes \hat{B}^{-}_{\alpha}+ \text{h.c.}$, where  $\mathbf{s}_{\alpha}$ is a spin residing at the $\alpha$th lattice site, with $\hat{s}^{\pm}_\alpha\equiv \hat{s}^{x}_\alpha\pm i\hat{s}^{y}_\alpha$,  and $\hat{B}^{\pm}_{\alpha}=\hat{B}_\alpha^x\pm i\hat{B}_\alpha^y$  is an operator acting on the reservoir.
Tracing out the degrees of freedom in the reservoir with the Born-Markov approximation \cite{breuer2002theory,rivas2012open,manzano2020short}, the Liouvillian master equation of the density operator $\rho$ of the spin chain can be written as
\begin{align}
    \frac{d\rho}{dt} = \mathcal{L}[\rho]=-i \left[ \mathcal{H}, \rho \right] + \mathcal{D}[\rho] \,.  \label{eq1}
\end{align}
Here the Hermitian Hamiltonian $\mathcal{H}$ and the Lindblad dissipator $\mathcal{D}$  describe, respectively, the coherent and dissipative time evolution as
\begin{align}
\mathcal{H}=&-\sum_{\alpha=1}^N\omega_{\alpha}\hat{s}^z_\alpha+\frac{1}{2}\sum_{\alpha,\beta=1\atop \alpha\neq\beta}^N
\mathcal{J}_{\alpha\beta}\hat{s}_{\alpha}^+ 
 \, \hat{s}_{\beta}^- \, \label{hamiltonian} , \\
\mathcal{D}[\rho] =&\sum_{\alpha,\beta=1}^N\Gamma_{\alpha \beta}\left(\hat{s}_\beta^+\rho \hat{s}_\alpha^--\frac{1}{2}\{\hat{s}_\alpha^-\hat{s}_\beta^+,\rho\}\right) \nonumber\\
+&\sum_{\alpha,\beta=1}^N\widetilde{\Gamma}_{\alpha \beta}\left(\hat{s}_\beta^-\rho \hat{s}_\alpha^+-\frac{1}{2}\{\hat{s}_\alpha^+\hat{s}_\beta^-,\rho\}\right) .
\label{eqlin}
\end{align}
Equations~(\ref{hamiltonian}) and (\ref{eqlin}) show that the coupling to the reservoir induces several effects: (i) a shift in the Zeeman frequency of individual spins, which is absorbed into $\omega_{\alpha}>0$; 
(ii) a coherent spin interaction $\mathcal{J}_{\alpha\beta}$ between the $\alpha$th and $\beta$th spin; (iii) local spin loss $\Gamma_{\alpha\alpha}$  and pump 
$\widetilde{\Gamma}_{\alpha\alpha}$;
and (iv) dissipative couplings in the form of correlated spin relaxation $\Gamma_{\alpha\beta}$ and excitation $\widetilde{\Gamma}_{\alpha\beta}$, where $\alpha\neq\beta$.
These parameters are determined by the dynamic correlation functions of the reservoir degrees of freedom. Therefore, $\Gamma_{\alpha\beta}$ and $\widetilde{\Gamma}_{\alpha\beta}$ are connected via detailed balance of the equilibrium reservoir. Assuming a reservoir with gapped bosonic excitations, here we neglect the Ising-type induced interactions and pure dephasing effects, as they arise from reservoir fluctuations at low frequencies~\cite{zou2022bell,li2025solid}, which are suppressed by the spectrum gap of the reservoir. 

In principle, the Liouvillian superoperator $\mathcal{L}$ can be fully diagonalized to yield its eigenspectrum and eigenmodes; the latter can exhibit localization at the boundaries, a phenomenon named the Liouvillian skin effect (LSE)\cite{Haga2021,yang2022liouvillian,mao2024liouvillian}. While this approach provides a complete description of the dissipative dynamics, a more analytically tractable perspective emerges by considering the conditional evolution of quantum trajectories in the absence of quantum jumps. Namely, 
upon the postselection of quantum trajectories that suppresses the quantum jump terms [those proportional to $\hat{s}^+_\beta\rho\hat{s}_\alpha^-$ in Eq.~(\ref{eqlin})], the conditional time evolution can be described by a non-Hermitian Hamiltonian $ \mathcal{H}_{\text{nh}}$ as:
\begin{align}
\frac{d\rho}{dt}=-i(\mathcal{H}_\text{nh}\rho-\rho \mathcal{H}_\text{nh}^\dag),  \label{deftrucatedh}
\end{align}
where
\begin{align}
    \mathcal{H}_{\text{nh}}
 &=-\text{Tr}[(\boldsymbol{\Omega}+i\widetilde{\boldsymbol{\Gamma}}_0)\boldsymbol{\mathcal{S}}^z]-\frac{i}{2} \, \mathbf{s}^{-} \boldsymbol{\mathcal{K}}_\text{nh} \, \left(\mathbf{s}^{+}\right)^T.
\label{effhamilexp}
\end{align}
The Hamiltonian~\eqref{effhamilexp} encapsulates the coherent and dissipative ingredients shaping the conditional dynamics, i.e.,  an effective complex Zeeman field, along with longitudinal and transverse spin-spin interaction terms.
 For brevity, a matrix form is used as follows: $\boldsymbol{\Omega} = \text{diag}(\omega_1, \ldots, \omega_N)$, $\widetilde{\boldsymbol{\Gamma}}_0=\text{diag}(\widetilde{\Gamma}_{11},\cdots,\widetilde{\Gamma}_{NN})$, $\boldsymbol{\mathcal{S}}^z = \text{diag}(\hat{s}_1^z, \ldots, \hat{s}_N^z)$,  $\mathbf{s}^{\pm} = (\hat{s}^{\pm}_1, \ldots, \hat{s}^{\pm}_N)$. Similarly, we adopt the notation of the interaction matrix $\boldsymbol{\mathcal{J}}$ in the Hermitian Hamiltonian (\ref{hamiltonian}) and the coefficient matrices $\boldsymbol{\Gamma}$ and $ \widetilde{\boldsymbol{\Gamma}}$ in the dissipator (\ref{eqlin})~\cite{supp}. The transverse coupling matrix reads as
\begin{equation}
    \boldsymbol{\mathcal{K}}_\text{nh} = i \boldsymbol{\mathcal{J}}^* + \boldsymbol{\Gamma} + \widetilde{\boldsymbol{\Gamma}}^*.
\end{equation}
The conditional Hamiltonian \eqref{effhamilexp} can exhibit the NHSE, whose origin is rooted in the topological structure of non-Hermitian band theory~\cite{bergholtz2021exceptional,okuma2023non}.
Within our framework, the transverse coupling terms in the non-Hermitian spin Hamiltonian govern the propagation of spin excitations between different sites. This structure offers a natural interpretation of the NHSE in terms of nonreciprocal spin transport—mirroring the asymmetric hopping processes central to the Hatano-Nelson model. Namely, under open boundary conditions, an imbalance in the coupling strengths, such as $|\left(\mathcal{K}_{\text{nh}}\right)_{\alpha, \alpha+1}| > |\left(\mathcal{K}_{\text{nh}}\right)_{\alpha+1, \alpha}|$, leads to a nonreciprocal flow of spin excitations and their accumulation at one boundary of the chain. This directional transport may be further modulated by spin loss and longitudinal interactions~\cite{mao2023,li2024}, which can imprint observable signatures of the NHSE in the steady-state spin profile.

In light of the connection between NHSE and nonreciprocal transport, we return to the master equation~(\ref{eq1}) and examine the
time evolution of the two-point correlation functions under the full Liouvillian. The equations of motion can be put into a compact form as follows:
\begin{align}
    \frac{d}{dt}\langle\boldsymbol{\mathcal{A}}\rangle
    = - \langle \boldsymbol{\mathcal{N}}\boldsymbol{\mathcal{A}}\rangle
    - \langle \boldsymbol{\mathcal{A}}\boldsymbol{\mathcal{N}}^\dagger\rangle +4\langle\boldsymbol{\mathcal{S}}^z\widetilde{\boldsymbol{\Gamma}}\boldsymbol{\mathcal{S}}^z\rangle,\label{dymatrixspin}
\end{align}
where $\mathcal{A}_{\alpha \beta}= \hat{s}^-_\alpha \hat{s}^+_\beta$, $\boldsymbol{\mathcal{N}} =-i\boldsymbol{\Omega} + \boldsymbol{\Gamma}_0 + \boldsymbol{\mathcal{S}}^z\boldsymbol{\mathcal{K}}_\mathcal{L}  $ is the  the dynamical matrix, $\boldsymbol{\Gamma}_0=\text{diag}(\Gamma_{11},\dots,\Gamma_{NN})$, and  $\boldsymbol{\mathcal{K}}_\mathcal{L} = -i\boldsymbol{\mathcal{J}} + \boldsymbol{\Gamma}^* - \widetilde{\boldsymbol{\Gamma}}$~\cite{supp}. For $\alpha=\beta$, Eq.(\ref{dymatrixspin}) yields the time evolution of the density distribution of the spin excitations along the chain. Despite the difficulty in solving these dynamical equations, a key observation is $\boldsymbol{\mathcal{K}}_\mathcal{L}$ =  $\boldsymbol{\mathcal{K}}_\text{nh}^*$ when $\widetilde{\boldsymbol{\Gamma}} = 0$, indicating that LSE and NHSE may be consistent from the point of view of nonreciprocal spin transport under the condition of vanishing $\widetilde{\boldsymbol{\Gamma}}$. This condition can be satisfied by going to the zero-temperature limit of the non-driven reservoir, where the correlated spin excitation processes are fully suppressed. In what follows, we set the reservoir at zero temperature, hence $\widetilde{\boldsymbol{\Gamma}} = 0$, and further address the consistency between LSE and NHSE in the dilute magnon regime.

\section{Dilute magnon regime}

We proceed to focus on an array of large spins, i.e., $s \gg 1$, and employ the semiclassical Holstein-Primakoff transformation, i.e., $\hat{s}_{\alpha}^+ \approx \sqrt{2s}\, \hat{a}_{\alpha}$ and $\hat{s}_{\alpha}^z = s - \hat{a}_{\alpha}^\dag \hat{a}_{\alpha}$, where  $\hat{a}_{\alpha}$ ($\hat{a}_{\alpha}^\dag$) is the magnon annihilation (creation) operator. Within the regime of low magnon density, i.e., $n_\alpha = \langle \hat{a}_{\alpha}^\dag \hat{a}_{\alpha} \rangle \ll s$,  we neglect magnon-magnon interactions and truncate both the non-Hermitian Hamiltonian and the Lindbladian to quadratic order in magnon operators. 
Equation~\eqref{effhamilexp} takes then  the form of a Hatano-Nelson Hamiltonian~\cite{hatano1996localization,okuma2023non}:
\begin{align}
    \mathcal{H}_{\text{m}}
=\boldsymbol{\hat{a}}^\dag(-i\boldsymbol{\mathcal{H}}_{\text{m}})\boldsymbol{\hat{a}}^T,\label{magnonham}
\end{align}
where $\boldsymbol{\hat{a}}=(\hat{a}_1,\cdots,\hat{a}_N)$, and $\boldsymbol{\mathcal{H}}_{\text{m}}=i\mathbf{\Omega}+ s(i \boldsymbol{\mathcal{J}}^* +  \boldsymbol{\Gamma})$~\cite{supp}.

At the same time, we obtain the Lindbladian superoperator~\eqref{eq1} for noninteracting magnons, which can be solved exactly with the third quantization method~\cite{prosen2008third,prosen2010quantization,barthel2022solving,mcdonald2023third}, as
\begin{align}
    {\mathcal{L}}_m
     &=\boldsymbol{\hat{b}}
     \left(
     \begin{array}{cc}
         0 & -\mathbf{X}^T \\
       - \mathbf{X}  & 0
     \end{array}
     \right)\boldsymbol{\hat{b}}^T+s\,\text{Tr}\boldsymbol{\Gamma},\label{magnonliouvillian}
\end{align}
with
\begin{align}
    \mathbf{X}=\frac{1}{2}\left(
    \begin{array}{cc}
       \boldsymbol{\mathcal{H}}_{m}  & 0 \\
       0  & \boldsymbol{\mathcal{H}}_{m}^*
    \end{array}
    \right),\label{matx}
\end{align}
where $\boldsymbol{\hat{b}}
= (\boldsymbol{\hat{a}}^L, \boldsymbol{\hat{a}}^{\dag R}, \boldsymbol{\hat{a}}^{\dag L} - \boldsymbol{\hat{a}}^{\dag R}, \boldsymbol{\hat{a}}^R - \boldsymbol{\hat{a}}^L)$, and the superscripts $R$ and $L$ indicate that the operator acts on the vectorized density matrix $\rho$ from the right and left sides, respectively~\cite{supp}. It is clear that the eigenenergy spectrum and eigenmodes of $\mathcal{L}_m$ are fully determined by the  non-Hermitian Hamiltonian~(\ref{magnonham}). Thus, the system exhibits NHSE and LSE at the same level in the low-excitation and low-temperature limit.

This equivalence is further substantiated by examining the transport properties of the system.  As a result of neglecting magnon interactions, the equations of motion governed by the Liouvillian (\ref{magnonliouvillian}) are closed in the quadratic order,   and coincide with those obtained from the non-Hermitian formalism:
\begin{align}
    \frac{d }{dt}\boldsymbol{\mathcal{C}}
    &=-\boldsymbol{\mathcal{H}}^*_{\text{m}}\boldsymbol{\mathcal{C}}-\boldsymbol{\mathcal{C}}(\boldsymbol{\mathcal{H}}_m^*)^\dag,\label{dyeqmagnon}
\end{align}
where $\mathcal{C}_{\alpha\beta} = \langle \hat{a}^\dag_{\alpha} \hat{a}_{\beta} \rangle$ is the magnon two-point correlation function~\cite{supp}. Focusing on the magnon density distribution $n_\alpha=\langle \hat{a}^\dag_\alpha \hat{a}_\alpha \rangle$, it becomes clear that a  magnon accumulation at an open boundary can emerge when the hopping along the chain is nonreciprocal, which is allowed by the non-Hermitian nature of the hopping matrix $\boldsymbol{\mathcal{H}}_{\text{m}}$.

In summary, the non-Hermitian Hamiltonian captures not only the spectral features associated with NHSE but also the steady-state and dynamical manifestations of LSE and nonreciprocal spin transport in the dilute magnon and zero-temperature regime. Nonetheless, beyond this regime, the non-Hermitian Hamiltonian description can deviate from the full Liouvillian dynamics.

\begin{figure*}[]
    \centering
\includegraphics[width=0.95\linewidth]{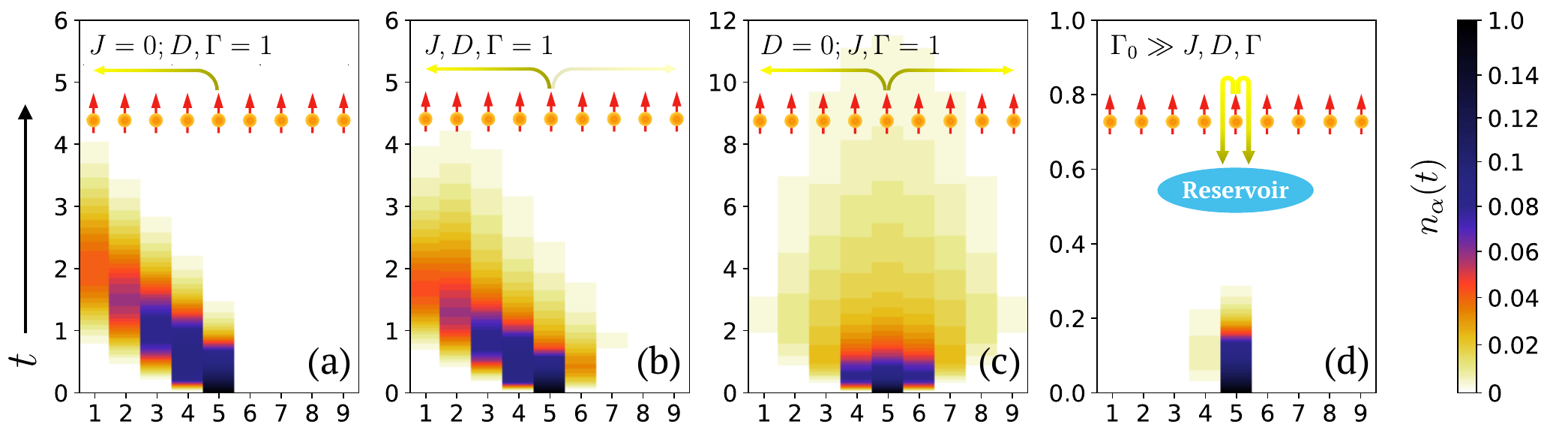}  
\label{fig:dynamics1}
    \caption{(a)-(d) The dynamical evolution of the magnon number $n_{\alpha}$ as a function of time. At time $t=0$, a magnon resides at the site $\alpha=5$. (a) For $J=0$ and $D,\Gamma=1$ the magnonic excitation propagates only towards the left of the array.  (b) Nonreciprocal magnon propagation towards the left  and right sides of the array for $J,D,\Gamma=1$. (c) For $D=0$ and $\Gamma, J=1$, the propagation is reciprocal. In (a)-(c), the local dissipation is set to $\Gamma_{0}= 2\Gamma$. (d) For a local dissipation $\Gamma_{0} \gg D, \Gamma \neq 0$,  magnon decay can suppress the spreading, such that no NHSE is observable.
    } 
\label{fig:dynamics}
\end{figure*}

\section{Nonreciprocity}
To explore further the role of the NHSE and LSE on magnon dynamics, we focus a spin chain with nearest-neighbor couplings, uniform on-site Zeeman frequency ($\omega_\alpha \equiv \omega$), and local dissipation ($\Gamma_{\alpha\alpha} \equiv \Gamma_0$). For simplicity, we restrict the nonlocal dissipation to its symmetric component, defining $\Gamma = \text{Re}[\Gamma_{\alpha,\alpha+1}] = \text{Re}[\Gamma_{\alpha+1,\alpha}]$. The nearest-neighbor coherent interaction is taken as $\mathcal{J}_{\alpha,\alpha+1} = (\mathcal{J}_{\alpha+1,\alpha})^* = J + iD$, where $J = \text{Re}[\mathcal{J}_{\alpha,\alpha+1}]$ and $D = \text{Im}[\mathcal{J}_{\alpha,\alpha+1}]$ correspond to the symmetric and antisymmetric exchange terms, respectively. Within this framework, the non-Hermitian magnon Hamiltonian~\eqref{magnonham} becomes:
\begin{align}
\mathcal{H}_\text{m}=\mathcal{H}^0_\text{m}+
\sum_{\alpha=1}^{N-1}
\left(\gamma_R\hat{a}_{\alpha+1}^\dag \hat{a}_{\alpha} + \gamma_L\hat{a}_{\alpha}^\dag \hat{a}_{\alpha+1}\right)\,.\label{magonequ}
\end{align}
Here, $\mathcal{H}^0_\text{m} = \epsilon_0\sum_{\alpha=1}^N \hat{a}^\dag_\alpha \hat{a}_\alpha$, with $\epsilon_0=\omega - is\Gamma_0$, represents an effective complex on-site potential, while $\gamma_R = s[J + i(D - \Gamma)]$ and $\gamma_L = s[J - i(D + \Gamma)]$ denote the rightward and leftward hopping amplitudes, respectively. When both $D$ and $\Gamma$ are nonzero, these amplitudes are asymmetric, resulting in $|\gamma_R / \gamma_L| \neq 1$. 
Under open boundary conditions, this asymmetry induces localization of the eigenmodes that characterizes the NHSE. We confirm this by diagonalizing Eq.(\ref{magonequ}), which yields 
\begin{align}
\left|\psi^{R}_{n,\alpha}\right|^2 = \left|\gamma_R/\gamma_L\right|^{ \alpha} \sin^2\left( \frac{n\alpha\pi}{N+1} \right),
\label{285}
\end{align}
for the probability density distributions of the right eigenvectors,  with \(n = 1, \ldots, N\)~\cite{supp}. Equation~\eqref{285} reveals a generic accumulation of eigenstates at the system boundaries [see Fig.~\ref{fig:master}(b)], with the direction of localization controlled by the hopping asymmetry.

Under periodic boundary conditions, the non-Hermitian Hamiltonian has the momentum-space form, $\hat{\mathcal{H}}_\text{m}=\sum_{k}(\epsilon_0+\epsilon_k)\hat{a}^\dag_k\hat{a}_k$, where $\epsilon_k=2s[\left(J\cos k + D\sin k \right)-i\Gamma\cos k ]$ is the eigenenergy spectrum.  As shown in Fig.~\ref{fig:master} (c)$, \epsilon_{k}$ has a point gap in the complex plane. The winding number of the spectrum around this point gap underpins the  topological origin of NHSE~\cite{okuma2020topological,gong2018,bergholtz2021exceptional,Brunelli2023}. In the limiting case where either $\gamma_L$ or $\gamma_R$ vanishes, the system enters a unidirectional hopping regime. Under this condition, all eigenstates become exponentially localized at a single boundary—either the left or the right edge—depending on the directionality of the hopping. Correspondingly, the complex energy spectrum traces a circle in the complex plane.

Even in this case, however, it is not clear whether the boundary localization of eigenstates and the associated finite spectral winding number necessarily lead to observable signatures of nonreciprocal magnon transport. To address this question, we simulate the real-time evolution of the magnon number $n_\alpha(t)$ by numerically solving 
 Eq.~(\ref{dyeqmagnon}) for an initial state with a single magnon excitation at the center of a $N=9$ spin array, i.e., setting $ \langle\hat{a}^\dag_{\alpha}\hat{a}_{\beta}\rangle=0$ for all except for $\alpha = \beta =5$. 
The resulting dynamics of the magnon number $n_\alpha(t)$ are shown in Fig.~\ref{fig:dynamics}(a)-(c) for
unidirectional, nonreciprocal, and reciprocal propagation, respectively. 
Away from the initial excitation site, the buildup of magnon density competes with dissipation, resulting in finite time and length scales over which the NHSE can be dynamically observed.  Such scales are determined jointly by the hopping amplitudes and the local dissipation, and are not captured by the non-Hermitian analysis alone~\cite{bergholtz2021exceptional,borgnia2020non,okuma2020topological,lee2023anomalously}.
As illustrated by Fig.~\ref{fig:dynamics}(d), a large local dissipation can fully suppress the state accumulation at the system boundary. 
The observation of the excitations accumulating at the edge can be facilitated by turning on the spin pumping.  However, as our previous analysis demonstrates, this driven-dissipative regime lies beyond the scope of the non-Hermitian Hamiltonian description and requires a full treatment within the Lindblad formalism to accurately capture the steady-state behavior.

Our analysis also reveals that the key ingredient for nonreciprocity is the balance between DMI-like exchange interactions (\(\propto D\)) and nonlocal, spatially symmetric dissipation (\(\propto \Gamma\)), in close analogy with the reservoir-engineering approaches proposed in Refs.~\cite{metelmann2015nonreciprocal,fang2017generalized} for realizing nonreciprocal photonic devices.
In the context of our model, a finite coherent chiral interaction $D_{\alpha\beta}=\text{Im}[\mathcal{J}_{\alpha\beta}]$ requires the bath to break both time-reversal and inversion symmetries~\cite{li2024cooperative}. Such scheme might be effectively realized using noncentrosymmetric magnetic or chiral phononic baths~\cite{flebus2023phonon}

However, these ingredients can also naturally arise -- beyond reservoir engineering setups --  in magnetic crystal that display intrinsic DMI interactions and non-local reciprocal dissipation due to interactions with the surrounding crystalline environment~\cite{reyes2024nonlocal}.
 This latter case is analogous to the model discussed by Deng and coauthors~\cite{deng2022non} in a 2$d$ layer, which identified DMI and nonlocal dissipation as the necessary ingredients to generate the finite spectral area that signals the emergence of the NHSE in this higher dimension~\cite{zhang2022universal}.  
 Magnetic heterostructures provide also a natural platform in which both effects coexist, as nonmagnetic spacer layers can mediate DMI and nonlocal spin pumping~\cite{kim2013chirality,shen2022effects,camley2023consequences,pogoryelov2020nonreciprocal,tserkovnyak2005nonlocal,tserkovnyak2002spin}.
We explore the connection between their magnetization dynamics and our microscopic model  in the following section.

\section{Magnetization dynamics} 
\begin{figure}[t]
    \centering
\includegraphics[width=0.95\linewidth]{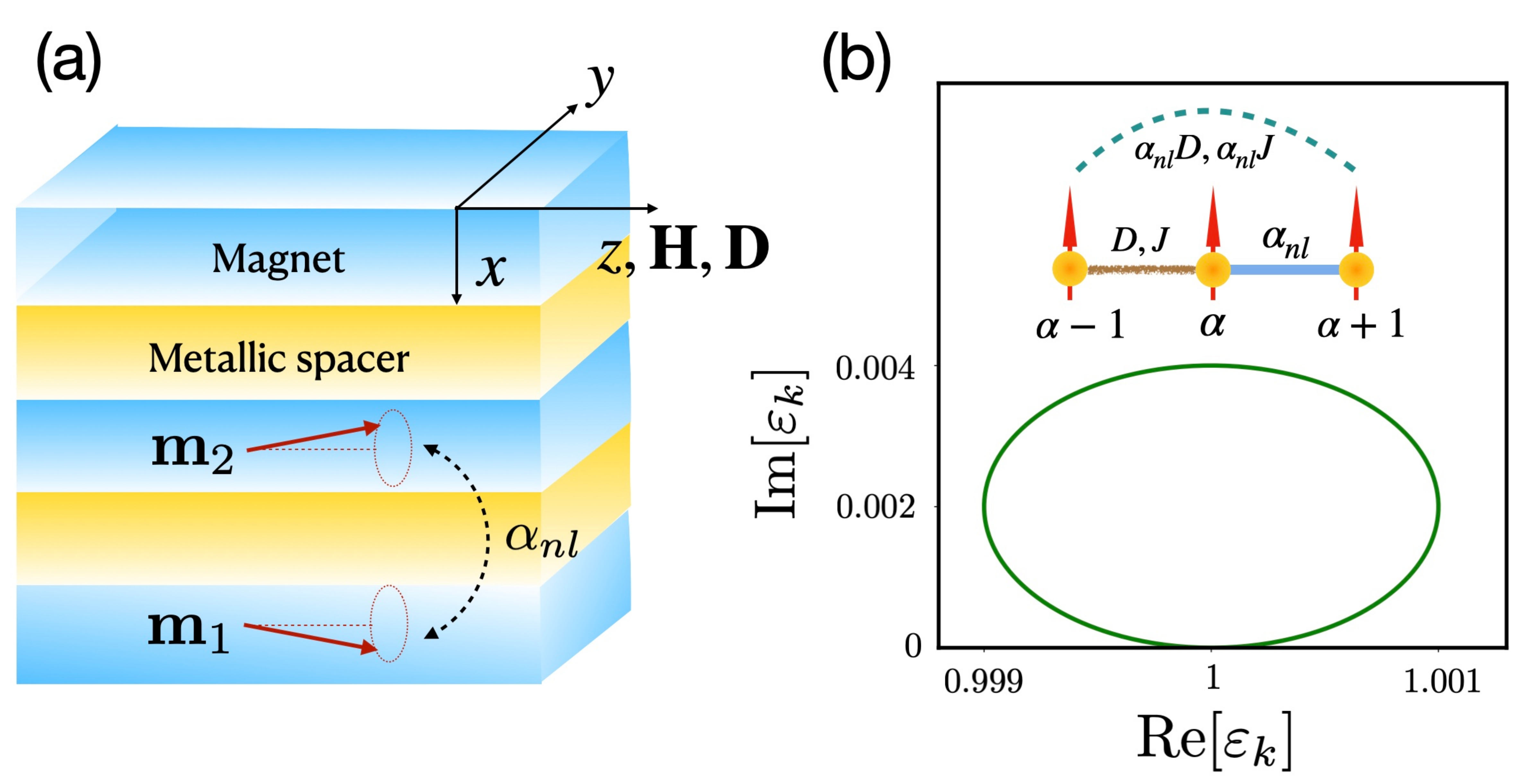}  
    \caption{(a) Schematic of a multilayer magnetic array.  The metallic spacers can mediate an interlayer DM interaction $\propto D$ and a damping-like spin pumping torque $\propto \alpha_{nl}$ between nearest neighbor magnetic layers. (b) Elliptical eigenenergy loop  in the complex plane.   
    Small effective next nearest neighbor interactions $\propto \alpha_{nl} D, \alpha_{nl} J$ are induced dynamically and preclude 
    fully circular energy loop, even when the nearest neighbor hoppings become unidirectional. We plot the elliptical eigenenergy loop by choosing the local dissipation as $\alpha_{l}=0.002$ and the nonlocal dissipation as $\alpha_{nl}=0.001$. 
    } 
\label{fig:LLG}
\end{figure}

We take as a case study the magnetic multilayer sketched in Fig.~\ref{fig:LLG}(a), where each magnetic layer interacts with its nearest neighbors via a metallic spacer. In addition to the intrinsic Gilbert damping $\alpha_{l}$ of the magnetic dynamics, the metallic spacer mediates a nonlocal spin pumping $\alpha_{nl}$ between the long-wavelength magnetization dynamics of adjacent layers~\cite{arxivSklenar,PhysRevB.66.224403,RevModPhys.77.1375,PhysRevLett.88.117601}. 
 The electric Fermi surface in metallic spacer can also mediates an effective coherent RKKY coupling $J$, with a DMI component $D$ due to interfacial inversion-symmetry  breaking~\cite{huang2022growth,di2015direct,ma2020longitudinal,han2019long,fernandez2019symmetry,avci2021chiral} .  

A minimal model for the magnetic Hamiltonian of the  multilayer can be written as~\cite{yuan2023unidirectional}
$\tilde{\mathcal{H}}= -\sum_{\langle \alpha\beta\rangle}
   \left[ J\mathbf{m}_\alpha\cdot\mathbf{m}_\beta 
   +D \hat{\mathbf{z}}\cdot(\mathbf{m}_\alpha\times\mathbf{m}_\beta) \right] 
   -\sum_{\alpha} \mu_0M_s\mathbf{H}\cdot \mathbf{m}_\alpha$
where $\mathbf{H}=H\hat{\mathbf{z}}$ is an externally applied magnetic field oriented along the $\hat{\mathbf{z}}$ direction, $M_s$ the saturation magnetization and $\mu_{0}$ the vacuum permeability. 
The macrospin dynamics of the magnetization  $\mathbf{m}_{\alpha}$ of the $\alpha$th layer is determined by the a modified LLG equation
 \begin{align}
 \frac{\partial \mathbf{m}_\alpha}{\partial t} &= -\frac{\gamma}{M_s}\mathbf{m}_{\alpha}\times\mathbf{H}_{\text{eff},\alpha}+\alpha_l \mathbf{m}_\alpha\times\frac{\partial\mathbf{m}_\alpha}{\partial t}\nonumber \\
 &+ \alpha_{nl} \mathbf{m}_\alpha\times \left(\frac{\partial\mathbf{m}_{\alpha-1}}{\partial t}+\frac{\partial\mathbf{m}_{\alpha+1}}{\partial t} \right)\,,
 \label{eq7}
 \end{align}
with $\mathbf{H}_{\text{eff},\alpha}=-\delta \tilde{\mathcal{H}}/\delta\mathbf{m}_\alpha $ and $\gamma$ the gyromagnetic ratio.
We linearize the LLG equation and derive its  spectrum under periodic boundary condition~\cite{supp}. As shown in Fig.~\ref{fig:LLG}(b), it has a point gap, singling the existence of the NHSE.

For a bilayer, a balance between DMI and nonlocal damping, i.e., $D=\pm \alpha_{nl}\mu_0M_sH$ can yield unidirectional transport~\cite{yuan2023unidirectional}.    
However, a key difference between the non-Hermitian model~(\ref{effhamilexp}) and Eq.~(\ref{eq7}) arises for more than two layers: the latter cannot exhibit exact unidirectionality.  
The nonlocal damping in Eq.~(\ref{eq7}) is determined for a given layer by the instantaneous dynamic state of the adjacent layers. This
effectively establishes next-to-nearest neighbor and even further interactions,
which stay active as the nearest-neighbor hopping in one direction can be turned off. 
Therefore, the dynamics can be nonreciprocal but not unidirectional, as reflected by the nonvanishing ellipticity of the eigenenergy loop. Nonetheless, despite the absence of strict unidirectionality, our results demonstrate that these platforms offer a promising avenue for the experimental realization of non-Hermitian skin effects.

\section{Discussion} 

In this work, we explore the emergence of the NHSE and its Liouvillian counterpart (LSE) in magnetic systems. By systematically bridging the non-Hermitian Hamiltonian framework with the full Lindblad dynamics, we identify the precise conditions—namely, the dilute magnon regime and absence of external pumping—under which the two frameworks become equivalent, with the NHSE and LSE necessarily implying one another. At the same time, we expose the limitations of the non-Hermitian Hamiltonian approach, which, despite capturing boundary-mode accumulation, fails to account for key dynamical features such as finite-time transport signatures and the role of strong local dissipation. We also identify the interplay between coherent chiral interactions (e.g., DMI) and nonlocal dissipative couplings as the minimal ingredients for realizing the NHSE, irrespective of whether they originate from engineered reservoirs or intrinsic crystalline environments. Finally, by mapping our microscopic model onto the LLG dynamics of experimentally accessible magnetic heterostructures, we establish these platforms as promising candidates for the observation and control of the NHSE.

\section{Acknowledgments.}
The authors thank R. A. Duine, J. Marino and V. Flynn for helpful discussions. 
This work was supported by the National Science Foundation under Grant No. 2337713.

\bibliography{library}

\end{document}


\title{Supplemental material:  ``Non-Hermitian  and Liouvillian skin effects in magnetic systems"}

\author{Xin Li}
\email{licqp@bc.edu}
\affiliation{Department of Physics, Boston College, 140 Commonwealth Avenue Chestnut Hill, MA 02467, USA}
\author{Mohamed Al Begaowe}
\affiliation{Department of Physics, Boston College, 140 Commonwealth Avenue Chestnut Hill, MA 02467, USA}
\author{Shu Zhang}
\affiliation{Collective Dynamics and Quantum Transport Unit, Okinawa Institute of Science and Technology Graduate University, Okinawa 904-0495, Japan}
\affiliation{Max Planck Institute for the Physics of Complex Systems, 01187 Dresden, Germany}
\author{Benedetta Flebus}

\affiliation{Department of Physics, Boston College, 140 Commonwealth Avenue Chestnut Hill, MA 02467, USA}

\date{\today}
\maketitle

\section*{S1. Master equation framework}

\subsection{Conditional non-Hermitian Hamiltonian for spin waves }

In this section, we derive the conditional non-Hermitian Hamiltonian (5) presented in the main text. We begin by rewriting the master equation (1) as
\begin{align}
    \frac{d}{dt}\rho
    &=-i[\mathcal{H},\rho]+\sum_{\alpha,\beta=1}^N  \Gamma_{\alpha \beta}\left(\hat{s}_\beta^+\rho \hat{s}_\alpha^--\frac{1}{2}\{\hat{s}_\alpha^-\hat{s}_\beta^+,\rho\}\right)+\sum_{\alpha,\beta=1}^N\tilde{\Gamma}_{\alpha \beta}\left(\hat{s}_\beta^-\rho \hat{s}_\alpha^+- \frac{1}{2}\{\hat{s}_\alpha^+\hat{s}_\beta^-,\rho\}\right),\nonumber\\
    &=-i\left(\mathcal{H}_{\text{nh}}\rho-\rho{\mathcal{H}}_{\text{nh}}^{\dag}\right)+\sum_{\alpha,\beta=1}^N  \Gamma_{\alpha \beta}\hat{s}_\beta^+\rho \hat{s}_\alpha^-+\sum_{\alpha,\beta=1}^N\tilde{\Gamma}_{\alpha\beta}\hat{s}_\beta^-\rho\hat{s}_\alpha^+.\label{effhamil}
\end{align}
By neglecting the last two terms in the second row of Eq.~(\ref{effhamil})—i.e., the quantum jump terms—we obtain a conditional evolution equation,

\begin{align}
    \frac{d}{dt}\rho
    &=-i\left(\mathcal{H}_{\text{nh}}\rho-\rho{\mathcal{H}}_{\text{nh}}^\dag\right),\label{trunceffhamil}
\end{align}
where the conditional non-Hermitian Hamiltonian $\mathcal{H}_{\text{nh}}$ is given as follows:
\begin{align}
\mathcal{H}_{\text{nh}}&=-\sum_{\alpha=1}^N\omega_\alpha \hat{s}^z_\alpha+\frac{1}{2}\sum_{\substack{\alpha,\beta=1\\ \alpha\neq\beta }}^N 
 \mathcal{J}_{\alpha\beta}\hat{s}_{\alpha}^+ 
 \, \hat{s}_{\beta}^--\frac{i}{2}\sum_{\alpha,\beta=1}^N\Gamma_{\alpha\beta}\hat{s}_\alpha^-\hat{s}_\beta^+-\frac{i}{2}\sum_{\alpha,\beta=1}^N\tilde{\Gamma}_{\alpha\beta}\hat{s}_\alpha^+\hat{s}_\beta^-,\nonumber\\
&=-\sum_{\alpha=1}^N\left(\omega_\alpha+i\tilde{\Gamma}_{\alpha\alpha}\right) \hat{s}^z_\alpha-\frac{i}{2}\sum_{\alpha,\beta=1}^N\left(i\mathcal{J}_{\beta\alpha}+\Gamma_{\alpha\beta}+\tilde{\Gamma}_{\beta\alpha}\right)\hat{s}_\alpha^-\hat{s}_\beta^+,\nonumber\\
&=-\text{Tr}[(\boldsymbol{\Omega}+i\boldsymbol{\tilde{\Gamma}}_0)\boldsymbol{\mathcal{S}}^z]-\frac{i}{2}\mathbf{s}^{-}\boldsymbol{\mathcal{K}}_{\text{nh}}(\mathbf{s}^{+})^T.\label{nonherhamspinwave}
\end{align}
Here we have used the commutation relation $[\hat{s}^+_{\alpha}, \hat{s}^-_{\beta}] = 2\delta_{\alpha\beta} \hat{s}^z_{\alpha}$ and the Hermiticity of the coherent interaction matrix $\boldsymbol{\mathcal{J}}$ and the decoherence matrix $\boldsymbol{\Gamma}$, i.e., $\mathcal{J}_{\beta\alpha} = \mathcal{J}^*_{\alpha\beta}$ and $\Gamma_{\beta\alpha} = \Gamma^*_{\alpha\beta}$. In the final row, we expressed $\mathcal{H}_{\text{nh}}$ in matrix form by introducing the vectors $\mathbf{s}^{-} = (\hat{s}_1^{-}, \cdots, \hat{s}_N^{-})$ and $\mathbf{s}^{+} = (\hat{s}_1^{+}, \cdots, \hat{s}_N^{+})^T$. Here, the matrices $\mathbf{\Omega}$, $\boldsymbol{\mathcal{S}}^z$, and $\boldsymbol{\mathcal{K}}_{\text{nh}}$ are defined as $\mathbf{\Omega} = \text{diag}(\omega_1, \cdots, \omega_N)$, $\boldsymbol{\mathcal{S}}^z = \text{diag}(s_1^z, \cdots, s_N^z)$, and $\boldsymbol{\mathcal{K}}_{\text{nh}} = i\boldsymbol{\mathcal{J}}^* + \boldsymbol{\Gamma} + \boldsymbol{\tilde{\Gamma}}^*$.

\subsection{Dynamical equations for two-point correlation functions}

We now  derive Eq.~(7) in the main text. We start by considering a general Lindblad master equation:
\begin{align}
    \frac{d}{dt}\rho=-i[\mathcal{H},\rho]+\mathcal{D}[\rho],
\end{align}
where the Lindbladian superoperator $\mathcal{D}[\rho]$ takes the form
\begin{align}
    \mathcal{D}[\rho]=\sum_{\alpha,\beta=1}^N\gamma_{\alpha\beta}\left(L_\beta\rho L_\alpha^\dag-\frac{1}{2}\{L_\alpha^\dag L_\beta,\rho\}\right),
\end{align}
with $L_{\alpha}$ and $L_{\beta}$ denoting the Lindblad operators. One can show that an operator $\mathcal{O}$ evolves according to the following equation: 

\begin{align}
    \frac{d}{dt} \mathcal{O}
    = i[\mathcal{H},\mathcal{O}]+\mathcal{D}^\dag[\mathcal{O}],\label{oopdjoint}
\end{align}
where the adjoint Lindbladian $\mathcal{D}^\dag[\mathcal{O}]$ is defined as
\begin{align}
    \mathcal{D}^\dag[\mathcal{O}]=\sum_{\alpha,\beta=1}^N\gamma_{\alpha\beta}\left(L_\alpha^\dag\mathcal{O}L_{\beta}-\frac{1}{2}\{L_{\alpha}^\dag L_{\beta},\mathcal{O}\}\right).
\end{align}
Taking $\mathcal{O}=\hat{s}^-_l\hat{s}^+_k$ and $L_\alpha=\hat{s}^-_\alpha$ in \eqref{oopdjoint}, we obtain the evolution equation for the two-point correlation operator $\hat{s}^-_l\hat{s}^+_k$ as follows:

\begin{align}
    \frac{d}{dt} \hat{s}^-_l\hat{s}^+_k =i [\mathcal{H},\hat{s}^-_l\hat{s}^+_k] + \mathcal{D}^\dag[\hat{s}^-_l\hat{s}^+_k].\label{twopointeq}
\end{align}
For the first term on the right hand side of Eq.~(\ref{twopointeq}), we have:

\begin{align}
    i[\mathcal{H},\hat{s}^-_l\hat{s}^+_k]&=i[-\sum_{\alpha=1}^N\omega_\alpha \hat{s}^z_\alpha+\frac{1}{2}\sum_{\substack{\alpha,\beta=1\\ \alpha\neq\beta }}^N 
 \mathcal{J}_{\alpha\beta}\hat{s}_{\alpha}^+ 
 \, \hat{s}_{\beta}^-,\hat{s}^-_l\hat{s}^+_k],\nonumber\\
 &=-i \hat{s}^-_l\hat{s}_{k}^+\omega_k-i\sum_{\substack{\alpha=1\\ \alpha\neq\,k}}^N\hat{s}^-_l\hat{s}_{\alpha}^+\mathcal{J}_{\alpha\,k}\hat{s}^z_k+i\omega_l\hat{s}^-_l\hat{s}^+_k+i\sum_{\substack{\alpha=1\\ \alpha\neq\,l}}^N\hat{s}^z_l\mathcal{J}_{l\alpha}\hat{s}_{\alpha}^-\hat{s}^+_k.\label{spinwaveham}
\end{align} 

For the second term on the right hand side of Eq.~(\ref{twopointeq}), we have:

\begin{align}
\mathcal{D}^\dag[\hat{s}^-_l\hat{s}^+_k] 
 &=\sum_{\alpha,\beta=1}^N  \Gamma_{\alpha \beta}\left(\hat{s}_\alpha^-\hat{s}^-_l\hat{s}^+_k \hat{s}_\beta^+-\frac{1}{2}\{\hat{s}_\alpha^-\hat{s}_\beta^+,\hat{s}^-_l\hat{s}^+_k\}\right) +\sum_{\alpha,\beta=1}^N\tilde{\Gamma}_{\alpha \beta}\left(\hat{s}_\alpha^+\hat{s}^-_l\hat{s}^+_k \hat{s}_\beta^-- \frac{1}{2}\{\hat{s}_\alpha^+\hat{s}_\beta^-,\hat{s}^-_l\hat{s}^+_k\}\right),\nonumber\\
  &=-\sum_{\alpha=1}^N  \Gamma_{\alpha \,l}\hat{s}_\alpha^-\hat{s}_l^z\hat{s}^+_k-\sum_{\alpha=1}^N  \Gamma_{k \alpha}\hat{s}^-_l\hat{s}^z_k\hat{s}_\alpha^+
 +\sum_{\alpha=1}^N\tilde{\Gamma}_{\alpha \,k}\hat{s}_\alpha^+\hat{s}^-_l\hat{s}^z_k+\sum_{\alpha=1}^N\tilde{\Gamma}_{l \alpha}\hat{s}^z_l\hat{s}^+_k\hat{s}_\alpha^-,\nonumber\\
 &=-\sum_{\alpha=1}^N  \hat{s}_l^z\Gamma^T_{l\alpha }\hat{s}_\alpha^-\hat{s}^+_k+\sum_{\alpha=1}^N\hat{s}^z_l\tilde{\Gamma}_{l \alpha}\hat{s}_\alpha^-\hat{s}^+_k-\sum_{\alpha=1}^N  \hat{s}^-_l\hat{s}_\alpha^+\Gamma^T_{\alpha\,k }\hat{s}^z_k
 +\sum_{\alpha=1}^N\hat{s}^-_l\hat{s}_\alpha^+\tilde{\Gamma}_{\alpha \,k}\hat{s}^z_k,\nonumber\\
 &\quad-  \Gamma_{l \,l}\hat{s}_l^-\hat{s}^+_k- \hat{s}^-_l\hat{s}^+_k\Gamma_{k k}+4\hat{s}^z_l\tilde{\Gamma}_{lk}\hat{s}^z_k,\label{spinwavelinb}
\end{align}
where we used the commutation relations, $[\hat{s}^+_\alpha,\hat{s}^-_\beta]=2\delta_{\alpha\beta}\hat{s}^z_\alpha$ and $[\hat{s}^z_\alpha,\hat{s}^{\pm}_\beta]=\pm\delta_{\alpha\beta}\hat{s}^{\pm}_\alpha$.

Combining~\eqref{spinwaveham} and \eqref{spinwavelinb}, we obtain
\begin{align}
    \frac{d}{dt}\hat{s}^-_l\hat{s}^+_k&=i\omega_ls^-_l\hat{s}^+_k-  \Gamma_{l \,l}\hat{s}_l^-\hat{s}^+_k+i\sum_{\substack{\alpha=1\\ \alpha\neq\,l}}^Ns^z_l\mathcal{J}_{l\alpha}\hat{s}_{\alpha}^-\hat{s}^+_k-\sum_{\alpha=1}^N  \hat{s}_l^z\Gamma^T_{l\alpha }\hat{s}_\alpha^-\hat{s}^+_k+\sum_{\alpha=1}^N\hat{s}^z_l\tilde{\Gamma}_{l \alpha}\hat{s}_\alpha^-\hat{s}^+_k\nonumber\\
    &\quad-i \hat{s}^-_l\hat{s}_{k}^+\omega_k- \hat{s}^-_l\hat{s}^+_k\Gamma_{k k}-i\sum_{\substack{\alpha=1\\ \alpha\neq\,k}}^N\hat{s}^-_l\hat{s}_{\alpha}^+\mathcal{J}_{\alpha\,k}\hat{s}^z_k-\sum_{\alpha=1}^N  \hat{s}^-_l\hat{s}_\alpha^+\Gamma^T_{\alpha\,k }\hat{s}^z_k
 +\sum_{\alpha=1}^N\hat{s}^-_l\hat{s}_\alpha^+\tilde{\Gamma}_{\alpha \,k}\hat{s}^z_k+4\hat{s}^z_l\tilde{\Gamma}_{lk}\hat{s}^z_k.
\end{align}
In terms of matrices, with $\mathcal{A}_{\alpha\beta}=\hat{s}^-_\alpha\hat{s}^+_\beta$, we finally obtain Eq.~(7) in the main text,
\begin{align}
    \frac{d}{dt}\langle\boldsymbol{\mathcal{A}}\rangle
    =-\langle\boldsymbol{\mathcal{N}}\boldsymbol{\mathcal{A}}\rangle-\langle\boldsymbol{\mathcal{A}}\boldsymbol{\mathcal{N}}^\dag\rangle+4\langle\boldsymbol{\mathcal{S}}^z\boldsymbol{\tilde{\Gamma}}\boldsymbol{\mathcal{S}}^z\rangle,
\end{align}
where the dynamical matrix $\boldsymbol{\mathcal{N}}$ is defined as
\begin{align}
    \boldsymbol{\mathcal{N}}=-i\mathbf{\Omega}+\boldsymbol{\Gamma}_0+\boldsymbol{\mathcal{S}}^z(-i\boldsymbol{\mathcal{J}}+\boldsymbol{\Gamma}^*-\boldsymbol{\tilde{\Gamma}})=-i\mathbf{\Omega}+\boldsymbol{\Gamma}_0+\boldsymbol{\mathcal{S}}^z\boldsymbol{\mathcal{K}}_{\mathcal{L}},
\end{align}
with 
\begin{align}
    \boldsymbol{\mathcal{K}}_{\mathcal{L}}=-i\boldsymbol{\mathcal{J}}+\boldsymbol{\Gamma}^*-\boldsymbol{\tilde{\Gamma}}.
\end{align}

\subsection{Magnon picture}

In the dilute magnon regime at zero temperature, we apply the Holstein–Primakoff transformation 
$\hat{s}^+ \approx \sqrt{2s}\,\hat{a}$, 
$\hat{s}^- \approx \sqrt{2s}\,\hat{a}^\dagger$, 
and $\hat{s}^z = s - \hat{a}^\dagger \hat{a}$ 
to express the conditional non-Hermitian spin Hamiltonian [Eq.~(8) of the main text] in terms of bosonic operators. 
Neglecting the constant energy shift $-s\sum_\alpha \omega_\alpha$, the resulting magnon Hamiltonian reads:

\begin{align}
    \mathcal{H}_{\text{m}}
    &=-\sum_{\alpha=1}^N\omega_{\alpha}(s-\hat{a}_\alpha^\dag\hat{a}_\alpha)+s\sum_{\substack{\alpha,\beta=1\\\alpha\neq\beta}}^N\mathcal{J}_{\alpha\beta}\hat{a}^\dag_{\beta}\hat{a}_{\alpha}-is\sum_{\alpha,\beta=1}^N\Gamma_{\alpha\beta}\hat{a}^\dag_{\alpha}\hat{a}_{\beta},\nonumber\\
&=\boldsymbol{\hat{a}}^\dag(\mathbf{\Omega}+s\boldsymbol{\mathcal{J}}^*-is\boldsymbol{\Gamma})\boldsymbol{\hat{a}}^T,\nonumber\\
&=\boldsymbol{\hat{a}}^\dag(-i\boldsymbol{\mathcal{H}}_{\text{m}})\boldsymbol{\hat{a}}^T,
\end{align}
where $\boldsymbol{\mathcal{H}}_{\text{m}} = i\mathbf{\Omega} + s(i\boldsymbol{\mathcal{J}}^* + \boldsymbol{\Gamma})$.

Next, we derive Eq.~(11) in the main text using Eq.~(\ref{oopdjoint}), i.e.,
\begin{align}
    \frac{d}{dt}\hat{a}_l^\dag\hat{a}_k&=i [\mathcal{H},\hat{a}_l^\dag\hat{a}_k] + \mathcal{D}^\dag[\hat{a}_l^\dag\hat{a}_k].\label{magnondynamical}
\end{align}

For the first term on the right-hand side of Eq.~\eqref{magnondynamical}, we obtain:
\begin{align}
    i [\mathcal{H},\,\hat{a}_l^\dag\hat{a}_k]&=i [-s\sum_{\alpha=1}^N\omega_\alpha+\sum_{\alpha=1}^N\,\omega_\alpha\hat{a}_\alpha^\dag\hat{a}_\alpha+s\sum_{\substack{\alpha,\beta=1\\\alpha\neq\beta}}^N 
 \mathcal{J}_{\beta\alpha}\hat{a}_{\alpha}^\dag\hat{a}_{\beta},\,\hat{a}_l^\dag\hat{a}_k],\nonumber\\
 &=-i \sum_{\alpha=1}^N \hat{a}_l^\dag\hat{a}_\alpha\omega_{\alpha}\delta_{\alpha\,k}+i \sum_{\alpha=1}^N\delta_{l\alpha}\omega_\alpha\,\hat{a}_\alpha^\dag\hat{a}_k-i s\sum_{\substack{\alpha=1\\\alpha\neq\,k}}^N 
 \mathcal{J}_{\alpha\,k}\hat{a}_l^\dag\,\hat{a}_{\alpha}+i s\sum_{\substack{\alpha=1\\\alpha\neq\,l}}^N 
 \mathcal{J}_{l\alpha}\hat{a}_{\alpha}^\dag\,\hat{a}_k.\label{cohemagon}
\end{align}

For the second term on the right-hand side of Eq.~\eqref{magnondynamical}, we find:

\begin{align}
    \mathcal{D}^\dag[\hat{a}_l^\dag\,\hat{a}_k]&=\sum_{\alpha,\beta=1}^N  2s\Gamma_{\alpha \beta}\left(\hat{a}_\alpha^\dag\,\hat{a}_l^\dag\,\hat{a}_k\,\hat{a}_{\beta} -\frac{1}{2}\{\hat{a}_\alpha^\dag\,\hat{a}_\beta,\hat{a}_l^\dag\,\hat{a}_k\}\right),\nonumber\\
    &= -s\sum_{\alpha=1}^N \Gamma^*_{l\alpha}\hat{a}_\alpha^\dag\,\hat{a}_k-s\sum_{\alpha=1}^N \Gamma_{k \alpha}\hat{a}_l^\dag\,\hat{a}_\alpha.
    \label{Lindmagon}
\end{align}

Combining Eqs. (\ref{cohemagon}) and (\ref{Lindmagon}), we obtain

\begin{align}
    \frac{d}{dt}\langle\hat{a}_l^\dag\hat{a}_k\rangle&=i\sum_{\alpha=1}^N\delta_{l\alpha}\omega_\alpha\langle\hat{a}_\alpha^\dag\hat{a}_k\rangle+i s\sum_{\substack{\alpha=1\\\alpha\neq\,l}}^N 
 \mathcal{J}_{l\alpha}\langle\hat{a}_\alpha^\dag\hat{a}_k\rangle-s\sum_{\alpha=1}^N \Gamma^*_{l\alpha}\langle\hat{a}_\alpha^\dag\,\hat{a}_k\rangle\nonumber\\
 &-i \sum_{\alpha=1}^N \langle\hat{a}_l^\dag\hat{a}_\alpha\rangle\omega_{\alpha}\delta_{\alpha\,k}-i s\sum_{\substack{\alpha=1\\\alpha\neq\,k}}^N 
 \langle\hat{a}_l^\dag\hat{a}_{\alpha}\rangle\mathcal{J}_{\alpha\,k}  -s\sum_{\alpha=1}^N \langle\hat{a}_l^\dag\hat{a}_\alpha\rangle\Gamma^*_{ \alpha\,k}.\label{magnoncorre}
\end{align}
In terms of matrix, Eq.~(\ref{magnoncorre}) can be rewritten as

\begin{align}
    \frac{d}{dt}\boldsymbol{\mathcal{C}}
    =-\boldsymbol{\mathcal{H}}_{\text{m}}^*\boldsymbol{\mathcal{C}}-\boldsymbol{\mathcal{C}}(\boldsymbol{\mathcal{H}}_{\text{m}}^*)^\dag,
\end{align}
which is Eq.~(11) in the main text.

\subsection{Third quantization method}

In this section, we apply the third quantization method~\cite{prosen2008third,prosen2010quantization} to derive Eq.~(9) of the main text. We start from the Lindbladian master equation expressed in terms of magnon operators:

\begin{align}
    \frac{d}{dt}\rho
    &=-i[\mathcal{H},\rho]+\mathcal{D}[\rho],
\end{align}
with
\begin{align}
    \mathcal{H}&=\,-s\sum_{\alpha=1}^N\omega_\alpha +\sum_{\alpha=1}^N\omega_\alpha \,\hat{a}_\alpha^\dag\,\hat{a}_\alpha+s\sum_{\substack{\alpha,\beta=1\\\alpha\neq\beta}}^N 
\mathcal{J}^T_{\beta\alpha}\hat{a}_{\beta}^\dag\,\hat{a}_{\alpha},\\
\mathcal{D}[\rho]
    &=\sum_{\alpha,\beta=1}^N  2s\Gamma_{\alpha \beta}\left(\hat{a}_{\beta}\rho \hat{a}_\alpha^\dag-\frac{1}{2}\{\hat{a}_\alpha^\dag\hat{a}_\beta,\rho\}\right).
\end{align}
The quantum jump terms,  such as \( \hat{a}_\alpha \rho \hat{a}_\beta^\dag \) and \( \hat{a}_\beta^\dag \rho \hat{a}_\alpha \),  represent left and right actions of operators on \( \rho \), with the first operator acting from the left and the second from the right. To formalize this structure, we introduce the left and right multiplication superoperators \( \hat{a}^L \) and \( \hat{a}^R \), defined by
\begin{align}
\hat{a}^L \ket{\rho} = \ket{\hat{a} \rho}, \qquad \hat{a}^R \ket{\rho} = \ket{\rho \hat{a}},
\label{super}
\end{align}
where \( \ket{\rho} \) denotes the vectorized form of the density matrix.
The superoperators~\ref{super} obey
the following relations
\begin{align}
    \hat{a}_\alpha^L\hat{a}_\beta^R\ket{\rho}=\ket{\hat{a}_\alpha\rho\hat{a}_\beta}=\hat{a}_\beta^R\hat{a}_\alpha^L\ket{\rho},\qquad [\hat{a}_\alpha^L,\hat{a}_\beta^R]=0.
\end{align}

Then the Hamiltonian part can be written as
\begin{align}
    -i[\mathcal{H},\rho]
    &\,\rightarrow\,(-i\mathcal{H}^L+i\mathcal{H}^R)\ket{\rho},\nonumber\\
&=\Big[-i\sum_{\alpha=1}^N\omega_\alpha (\hat{a}_\alpha^{\dag\,L}\hat{a}_\alpha^L- \hat{a}_\alpha^R\hat{a}_\alpha^{\dag\,R})+i\sum_{\substack{\alpha,\beta=1\\\alpha\neq\beta}}^N 
s\mathcal{J}^T_{\beta\alpha}(\hat{a}_{\alpha}^R\hat{a}_{\beta}^{\dag\,R}-\hat{a}_{\beta}^{\dag\,L}\hat{a}_{\alpha}^L)\Big]\ket{\rho}.
\end{align}

Note that the two operators $\hat{a}_\alpha^{\dag\,L}-\hat{a}_\alpha^{\dag\,R}$ and $\hat{a}_{\alpha}^R-\hat{a}_{\alpha}^L$ annihilate the identity operator $\hat{I}$,
\begin{align}
        (\hat{a}_\alpha^{\dag\,L}-\hat{a}_\alpha^{\dag\,R})\ket{\hat{I}}=\ket{(\hat{a}_\alpha^{\dag}-\hat{a}_\alpha^{\dag})}=0,\qquad (\hat{a}_\alpha^{R}-\hat{a}_\alpha^{L})\ket{\hat{I}}=\ket{(\hat{a}_\alpha-\hat{a}_\alpha)}=0,\qquad 
    \hat{a}^{L}\ket{\rho_0}=0,\qquad \hat{a}^{\dag\,R}\ket{\rho_0}=0,
\end{align}

which motivates the introduction of four new particle operators,

\begin{align}
    \hat{b}_{1,\alpha}=\hat{a}_{\alpha}^L,\quad \hat{b}_{2,\alpha}=\hat{a}_\alpha^{\dag\,R},\qquad  \hat{\tilde{b}}_{1,\alpha}=\hat{b}_{1,\alpha}^\dag-\hat{b}_{2,\alpha}=\hat{a}_\alpha^{\dag\,L}-\hat{a}_\alpha^{\dag\,R},\quad \hat{\tilde{b}}_{2,\alpha}=\hat{b}_{2,\alpha}^\dag-\hat{b}_{1,\alpha}=\hat{a}_{\alpha}^R-\hat{a}_\alpha^L,
\end{align}
satisfying the following commuting relation
\begin{align}
    \,[\hat{b}_{\mu,\alpha},\hat{b}_{\nu,\beta}]=0,\qquad [\hat{\tilde{b}}_{\mu,\alpha},\hat{\tilde{b}}_{\nu,\beta}]=0,\qquad [\hat{b}_{\mu,\alpha},\hat{\tilde{b}}_{\nu,\beta}]=\delta_{\mu\nu}\delta_{\alpha\beta}.
\end{align}

In terms of the new operators, the Hamiltonian part, can be written as

\begin{align}
    -i[\mathcal{H},\rho]
    &\,\rightarrow\,(-i\mathcal{H}^L+i\mathcal{H}^R)\ket{\rho},\nonumber\\
&=\left[-i\sum_{\alpha=1}^N\omega_\alpha (\hat{\tilde{b}}_{1,\alpha}\,\hat{b}_{1,\alpha}- \hat{\tilde{b}}_{2,\alpha}\,\hat{b}_{2,\alpha})+i\sum_{\substack{\alpha,\beta=1\\\alpha\neq\beta}}^N 
s\mathcal{J}^T_{\beta\alpha}(\hat{\tilde{b}}_{2,\alpha}\hat{b}_{2,\beta}-\hat{\tilde{b}}_{1,\beta}\,\hat{b}_{1,\alpha})\right]\ket{\rho},\nonumber\\
&=[-i(\hat{\boldsymbol{\tilde{b}}}_{1}\boldsymbol{\Omega}\,\hat{\boldsymbol{b}}_{1}^T- \hat{\boldsymbol{\tilde{b}}}_{2}\boldsymbol{\Omega}\,\hat{\boldsymbol{b}}_{2}^T)+i 
s(\hat{\boldsymbol{\tilde{b}}}_{2}\boldsymbol{\mathcal{J}}{\boldsymbol{b}}^T_{2}-\hat{\boldsymbol{\tilde{b}}}_{1}\boldsymbol{\mathcal{J}}^T\,\hat{\boldsymbol{b}}^T_{1})]\ket{\rho},\nonumber\\
&=[ {\hat{\boldsymbol{\tilde{b}}}_{1}}(-i\boldsymbol{\Omega}-i 
s\boldsymbol{\mathcal{J}}^*)\hat{\boldsymbol{b}}^T_{1}+\hat{\boldsymbol{\tilde{b}}}_{2}( i\boldsymbol{\Omega}+i 
s\boldsymbol{\mathcal{J}})\hat{\boldsymbol{b}}^T_{2}]\ket{\rho},\label{thirdcoherent}
\end{align}
with
\begin{align}
    \hat{\mathbf{b}}_{\mu}=(\hat{b}_{\mu,1},\cdots,\hat{b}_{\mu,N}),\qquad \hat{\boldsymbol{\tilde{b}}}_{\mu}=(\hat{\tilde{b}}_{\mu,1},\cdots,\hat{\tilde{b}}_{\mu,N}),\qquad \mu=1,2.
\end{align}

Similarly, we can rewrite the Lindbladian as 

\begin{align}
    \mathcal{D}[\rho]
    &\rightarrow\sum_{\alpha,\beta=1}^N 2s\Gamma_{\alpha \beta} \left(\hat{a}^L_{\beta} \hat{a}_{\alpha}^{\dag\,R}-\frac{1}{2}[(\hat{a}_\alpha^\dag\,\hat{a}_\beta)^L+(\hat{a}_\alpha^\dag\,\hat{a}_\beta)^R]\right)\ket{\rho},\nonumber\\
    &=\sum_{\alpha,\beta=1}^N -s\Gamma_{\alpha \beta} \left[(\hat{a}_\alpha^{\dag\,L}-\hat{a}_\alpha^{\dag\,R})\hat{a}^L_\beta+(\hat{a}_\beta^R-\hat{a}_\beta^L)\hat{a}_\alpha^{\dag\,R}\right]\ket{\rho},\nonumber\\
    &= -s\left(\hat{\boldsymbol{\tilde{b}}}_{1}\boldsymbol{\Gamma} \hat{\boldsymbol{b}}^T_{1}+\hat{\boldsymbol{\tilde{b}}}_{2}\boldsymbol{\Gamma}^T\hat{\boldsymbol{b}}^T_{2}\right)\ket{\rho}.\label{thirdlindbla}
\end{align}

Combining Eq.(\ref{thirdcoherent}) and (\ref{thirdlindbla}), we obtain 

\begin{align}
    \mathcal{L}_{\text{m}}
    &=- \hat{\boldsymbol{\tilde{b}}}_{1}[i\boldsymbol{\Omega}+s(i 
\boldsymbol{\mathcal{J}}^*+\boldsymbol{\Gamma})]\hat{\boldsymbol{b}}^T_{1}-\hat{\boldsymbol{\tilde{b}}}_{2}[-i\boldsymbol{\Omega}+s(-i 
\boldsymbol{\mathcal{J}}+\boldsymbol{\Gamma}^*)]\hat{\boldsymbol{b}}^T_{2},\nonumber\\
&=\hat{\boldsymbol{\tilde{b}}}_{1}(-\boldsymbol{\mathcal{H}}_{\text{m}})\hat{\boldsymbol{b}}^T_{1}+\hat{\boldsymbol{\tilde{b}}}_{2}(-\boldsymbol{\mathcal{H}}^*_{\text{m}})\hat{\boldsymbol{b}}^T_{2},\nonumber\\
&=\hat{\boldsymbol{\tilde{b}}}_{1}(-\frac{1}{2}\boldsymbol{\mathcal{H}}_{\text{m}})\hat{\boldsymbol{b}}^T_{1}+\hat{\boldsymbol{\tilde{b}}}_{2}(-\frac{1}{2}\boldsymbol{\mathcal{H}}^*_{\text{m}})\hat{\boldsymbol{b}}^T_{2}+\hat{\boldsymbol{\tilde{b}}}_{1}(-\frac{1}{2}\boldsymbol{\mathcal{H}}_{\text{m}})\hat{\boldsymbol{b}}^T_{1}+\hat{\boldsymbol{\tilde{b}}}_{2}(-\frac{1}{2}\boldsymbol{\mathcal{H}}_{\text{m}}^*)\hat{\boldsymbol{b}}^T_{2}\nonumber\\
&=\hat{\boldsymbol{\tilde{b}}}_{1}(-\frac{1}{2}\boldsymbol{\mathcal{H}}_{\text{m}})\hat{\boldsymbol{b}}^T_{1}+\hat{\boldsymbol{\tilde{b}}}_{2}(-\frac{1}{2}\boldsymbol{\mathcal{H}}_{\text{m}}^*)\boldsymbol{b}^T_{2}+\hat{\boldsymbol{b}}_{1}(-\frac{1}{2}\boldsymbol{\mathcal{H}}_{\text{m}})^T\hat{\boldsymbol{\tilde{b}}}^T_{1}+\hat{\boldsymbol{b}}_{2}(-\frac{1}{2}\boldsymbol{\mathcal{H}}_{\text{m}}^*)^T\hat{\boldsymbol{\tilde{b}}}^T_{2}+\frac{1}{2}\text{Tr}\boldsymbol{\mathcal{H}}_{\text{m}}+\frac{1}{2}\text{Tr}\boldsymbol{\mathcal{H}}^*_{\text{m}}.\label{linbdthirdfin}
\end{align}

Eq.~(\ref{linbdthirdfin}) can be rewritten in matrix form, 
\begin{align}  
\mathcal{L}_{\text{m}}
     &=(\hat{\boldsymbol{b}}_1,\hat{\boldsymbol{b}}_2,\hat{\tilde{\boldsymbol{b}}}_1,\hat{\tilde{\boldsymbol{b}}}_2)
     \left(
     \begin{array}{cc}
         0 & -\mathbf{X}^T \\
       - \mathbf{X}  & 0
     \end{array}
     \right)\left(
     \begin{array}{cccc}
     \hat{\boldsymbol{b}}_1\\
     \hat{\boldsymbol{b}}_2\\
     \hat{\boldsymbol{\tilde{b}}}_1\\
     \hat{\boldsymbol{\tilde{b}}}_2
     \end{array}
     \right)+s\text{Tr}\boldsymbol{\Gamma}\label{thirdlimcontex}
\end{align}
with
\begin{align}
    \mathbf{X}=\frac{1}{2}\left(
    \begin{array}{cc}
       \boldsymbol{\mathcal{H}}_{\text{m}}  & 0 \\
       0  & \boldsymbol{\mathcal{H}}_{\text{m}}^*
    \end{array}
    \right).\label{thirdmatcontex}
\end{align}
Eqs.~(\ref{thirdlimcontex}) and (\ref{thirdmatcontex}) are respectively Eqs.~(9) and (10) in the main text.

\subsection{Diagonalization of the effective non-Hermitian Hamiltonian for nearest neighbor interactions}

In the  non-unidirectional case, $\gamma_L\gamma_R\neq0$,   the non-Hermitian Hamiltonian in Eq.~(12) takes the form of a tridiagonal Toeplitz matrix:

\begin{eqnarray}
    \mathcal{H}_{\text{m}}=\left(
    \begin{array}{ccccccccc}
       \epsilon_0  & \gamma_L&&&& \\
        \gamma_R & \epsilon_0&\gamma_L&&&\\
        & \gamma_R&\ddots&\ddots&&\\
        &&\ddots & \ddots&\gamma_L&\\
        &&& \gamma_R& \epsilon_0&\gamma_L\\
        &&&&\gamma_R & \epsilon_0
    \end{array}
    \right)_{N\times N}.\label{nhhamilmat}
\end{eqnarray}
Diagonalizing Eq.~(\ref{nhhamilmat}), we obtain a series of eigenvalues:

 \begin{eqnarray}
    \lambda_n=\epsilon_0+2\sqrt{\gamma_L\gamma_R}\cos\frac{n\pi}{N+1},\quad n=1, \cdots, N\, ,
\end{eqnarray}
and the corresponding right and left eigenvectors 
\begin{eqnarray}
    |\Psi^R_n\rangle=(\psi^R_{n,1},\cdots,\psi^R_{n,\alpha},\cdots,\psi^R_{n,N})^T,\qquad
    |\Psi^L_n\rangle=(\psi^L_{n,1},\cdots,\psi^L_{n,\alpha},\cdots,\psi^L_{n,N})^T\,,
\end{eqnarray}
with
\begin{eqnarray}
    \psi^R_{n,\alpha}=\left(\frac{\gamma_R}{\gamma_L}\right)^{\alpha/2}\sin\frac{n\alpha\pi}{N+1}\,,\qquad   \psi^L_{n,\alpha}=\left(\frac{\gamma^*_L}{\gamma^*_R}\right)^{\alpha/2}\sin\frac{n\alpha\pi}{N+1}\,,\qquad\qquad n,\alpha =1,\cdots, N\,,
\end{eqnarray}
from which Eqs.~(13) of the main text follow directly.
On the other hand, if the system satisfies the unidirectional condition, i.e.,  $\gamma_L=0$ or $\gamma_R=0$, the Hamiltonian 
~(12) reduces to a Jordan  block of size $N$,

\begin{eqnarray}
\mathcal{H}^L_{\text{m}}=\left(
    \begin{array}{ccccccccc}
       \epsilon_0  & \gamma_L&&&& \\
        0 & \epsilon_0&\gamma_L&&&\\
        & 0&\ddots&\ddots&&\\
        &&\ddots & \ddots&\gamma_L&\\
        &&& 0& \epsilon_0&\gamma_L\\
        &&&&0 & \epsilon_0
    \end{array}
    \right)_{N\times N}
,\qquad\qquad
\mathcal{H}^R_{\text{m}}=\left(
    \begin{array}{ccccccccc}
       \epsilon_0  & 0&&&& \\
        \gamma_R & \epsilon_0&0&&&\\
        & \gamma_R&\ddots&\ddots&&\\
        &&\ddots & \ddots&0&\\
        &&& \gamma_R& \epsilon_0&0\\
        &&&&\gamma_R & \epsilon_0
    \end{array}
    \right)_{N\times N}.\label{Jordanblock}
\end{eqnarray}
The Jordan block Hamiltonians in Eq.~(\ref{Jordanblock}) are non-diagonalizable. However, one can immediately identify the single eigenvalue $\lambda=\epsilon_0$ with algebraic multiplicity $N$ and the corresponding right and left eigenvectors

\begin{eqnarray}
    |\Psi^{R}_L\rangle=(1,0,\cdots,0)^T,\qquad |\Psi^{L}_L\rangle=(0,\cdots,0,1)^T,\label{jordanupp}
\end{eqnarray}
for $\mathcal{H}^L_{\text{m}}$,  and

\begin{eqnarray}
    |\Psi^{R}_R\rangle=(0,\cdots,0,1)^T,\qquad |\Psi^{L}_R\rangle=(1,0,\cdots,0)^T,\label{jordandown}
\end{eqnarray}
for $\mathcal{H}^R_{\text{m}}$. Equations~(\ref{jordanupp}) and (\ref{jordandown}) demonstrate that, under unidirectional conditions, the eigenmodes become fully localized at one boundary, exemplifying the non-Hermitian skin effect.

\section{S2. Classical magnetization dynamics}

Equation (14) can be rewritten  as

\begin{eqnarray}
    \frac{\partial \mathbf{m}_\alpha}{\partial t} 
    &=& -\frac{\gamma J}{M_s}\mathbf{m}_{\alpha}\times(\mathbf{m}_{\alpha-1}+\mathbf{m}_{\alpha+1})+\frac{\gamma}{M_s}\mathbf{m}_{\alpha}\times[(\mathbf{m}_{\alpha+1}\times\mathbf{D})-(\mathbf{m}_{\alpha-1}\times\mathbf{D})] -\gamma\mu_0 \mathbf{m}_{\alpha}\times \mathbf{H}\nonumber\\
&&+\alpha_l\mathbf{m}_\alpha\times\frac{\partial\mathbf{m}_\alpha}{\partial t}+\alpha_{nl}\mathbf{m}_\alpha\times\frac{\partial\mathbf{m}_{\alpha-1}}{\partial t}+\alpha_{nl}\mathbf{m}_\alpha\times\frac{\partial\mathbf{m}_{\alpha+1}}{\partial t}\ . \label{CoupledLLG}
\end{eqnarray}
For small fluctuations of the magnetic order parameter around the equilibrium direction, i.e., $\mathbf{m}_\alpha=(m_{\alpha}^{x},m_{\alpha}^{y},1)$ with $|m_{\alpha}^{x(y)}|\ll1$, we can simplify Eq.~(\ref{CoupledLLG}) by only retaining terms linear in $m_{\alpha}^x$ and $m_{\alpha}^y$. Similarly, for  $\alpha_l,\alpha_{nl}\ll1$,   terms of second or higher order in these parameters can be neglected. Then, invoking the Holstein-Primakoff transformation,  i.e., $\langle\hat{a}_{\alpha}\rangle\equiv m_{\alpha}^x+ im_{\alpha}^y$, Eq.~(\ref{CoupledLLG}) can be written as
\begin{eqnarray}
\frac{\partial \hat{a}_\alpha}{\partial t} 
\approx &&+ \frac{\gamma}{M_s}\left\{\Big[(i-\alpha_l)\left(2 J +\mu_0 M_sH
\right)
+2 J\alpha_{nl}\right]\hat{a}_{\alpha}
+\Big[(-i+\alpha_l)\left( J -i D\right)-\alpha_{nl}\left(2 J 
+\mu_0 M_sH
\right)\Big]\hat{a}_{\alpha-1}\nonumber\\
&&+\Big[(-i+\alpha_l)\left( J+i D\right)-\alpha_{nl}\left(2 J+\mu_0M_sH\right)\Big]\hat{a}_{\alpha+1} +\alpha_{nl}\left( J-i D\right)\hat{a}_{\alpha-2} +\alpha_{nl}\left( J+i D\right)\hat{a}_{\alpha+2}\Big\} . \label{LLGlinear}
\end{eqnarray}
As shown by the last two terms on the right-hand-side of Eq.~(\ref{LLGlinear}), the nonlocal dissipation terms, i.e., $\alpha_{nl}\mathbf{m}_\alpha\times \left(\frac{\partial\mathbf{m}{\alpha-1}}{\partial t}+\frac{\partial\mathbf{m}{\alpha+1}}{\partial t} \right)$,  gives rise to effective next nearest neighbor  interactions. For a bilayer, i.e., $N=2$,  Eq.~(\ref{LLGlinear}) simplifies to

\begin{eqnarray}
\frac{\partial \hat{a}_1}{\partial t} 
&=&\frac{\gamma}{M_s}\Big\{\left[(i-\alpha_l)\left(2 J +\mu_0 M_sH
\right)
+2 J\alpha_{nl}\right]\hat{a}_{1}
+\Big[(-i+\alpha_l)\left( J+i D\right)-\alpha_{nl}\left(2 J+\mu_0M_sH\right)\Big]\hat{a}_{2}\Big\}\, ,\label{pairlone}\\
\frac{\partial \hat{a}_2}{\partial t} 
&=&\frac{\gamma}{M_s}\Big\{\left[(i-\alpha_l)\left(2 J +\mu_0 M_sH
\right)
+2 J\alpha_{nl}\right]\hat{a}_{2}
+\Big[(-i+\alpha_l)\left( J -i D\right)-\alpha_{nl}\left(2 J 
+\mu_0 M_sH
\right)\Big]\hat{a}_{1}\Big\}\ .  \label{pairltwo}
\end{eqnarray}
To achieve unidirectional transport, the second terms on the right-hand-side of Eqs.~(\ref{pairlone}) or (\ref{pairltwo}) must vanish, i.e.,

\begin{eqnarray}
    J=\pm\alpha_lD,\qquad D=\pm\alpha_{nl}\mu_0M_sH(1+\alpha^2_l-2\alpha_{nl}\alpha_l)^{-1}\, ,\label{relationfortwo}
\end{eqnarray}
where the positive and negative signs correspond to the unidirectionality of Eqs.~(\ref{pairlone}) and (\ref{pairltwo}), respectively.
Recalling the condition $\alpha_{nl}, \alpha_{l}\ll 1$,  Eqs.~(\ref{relationfortwo}) can be simplified to

\begin{eqnarray}
    J=0,\qquad D=\pm\alpha_{nl}\mu_0M_sH .\label{relationfortwosim}
\end{eqnarray}
However, for magnetic multilayers with $N\geq3$,  unidirectionality cannot be achieved due to the emergence of effective next nearest neighbor couplings. This can be shown explicitly by considering Eq.~\eqref{LLGlinear} in the limit of an infinitely-long spin chain, i.e., $N \rightarrow \infty$.
Invoking the Heisenberg equation of motion, i.e., $d\hat{a}/dt=-i[\hat{a},\mathcal{H}]$, and performing a Fourier transform, one can derive an effective Hamiltonian in momentum space, whose spectrum $\tilde{\varepsilon}_k$ reads as

\begin{eqnarray}
    \tilde{\varepsilon}_k=-\frac{\gamma}{M_s}\Big[1+i(\alpha_0+2\alpha_{nl}\cos k)\Big]\Big[ J(1-\cos k)+\mu_0 H M_s+ D\sin k\Big]\, .\label{mometumham}
\end{eqnarray}
Pluggling the unidirectionality condition~(\ref{relationfortwosim}) into Eq.~(\ref{mometumham}) yields

\begin{eqnarray}
    \tilde{\varepsilon}_k=-\gamma\mu_0H\Big[(1\pm\alpha_{nl}\sin k)+i(\alpha_l+2\alpha_{nl}\cos k)\Big]\, .\label{llgmomentum}
\end{eqnarray}
For simplicity, we normalize Eq.~(\ref{llgmomentum}) by dividing $-\gamma\mu_0H$ on both sides, i.e., $\varepsilon=\tilde{\varepsilon}_k/(-\gamma\mu_0H)$ . It is then straightforward to determine the relationship between the real and imaginary parts as

\begin{eqnarray}
    \frac{(\text{Re}\varepsilon_k-1)^2}{\alpha_{nl}^2}+\frac{(\text{Im}\varepsilon_k-\alpha_{l})^2}{4\alpha_{nl}^2}=1\, .\label{llgellipse}
\end{eqnarray}
One can readily recognize Eq.~(\ref{llgellipse}) as describing an ellipse rather than a circle, indicating the absence of unidirectionality but the presence of nonreciprocity.

\bibliographystyle{apsrev4-2}

\bibliography{suplibrary}